\documentclass[aps,prd,superscriptaddress
,twocolumn,twoside,
]{revtex4}

\usepackage{amsmath,graphicx,bm}
\usepackage{setspace,color}

\begin{document}

\title{On low-temperature structural phase transitions}

\author{A. Cano}
\email{andres@iaa.es}
\affiliation{\mbox{Departamento de F\'\i sica de la Materia Condensada C-III,
Universidad Aut\'onoma de Madrid, E-28049 Madrid, Spain}}
\affiliation{Instituto de Astrof\' \i sica de Andaluc\' \i a CSIC, PO Box 3004, E-18008 Granada, Spain}
\author{A. P. Levanyuk}
\email{levanyuk@uam.es}
\affiliation{\mbox{Departamento de F\'\i sica de la Materia Condensada C-III, 
Universidad Aut\'onoma de Madrid, E-28049 Madrid, Spain}}
\date{\today}

\begin{abstract}
We comment on zero- and low-temperature structural phase transitions, expecting that these comments might be relevant not only for this structural case. We first consider a textbook model whose classical version is the only model for which the Landau theory of phase transitions and the concept of ``soft mode'' introduced by Ginzburg are exact. Within this model, we reveal the effects of quantum fluctuations and thermal ones at low temperatures. To do so, the knowledge of the dynamics of the model is needed. However, as already was emphasized by Ginzburg {\it et al.} in eighties, a realistic theory for such a dynamics at high temperatures is lacking, what also seems to be the case in the low temperature regime. Consequently, some theoretical conclusions turn out to be dependent on the assumptions on this dynamics. We illustrate this point with the low-temperature phase diagram, and discuss some unexpected shortcomings of the continuous medium approaches. 
\end{abstract}

\maketitle

\section{Introduction}

Zero- and low-temperature ($T$) phase transitions are nowadays a subject of great interest (see, e.g., Refs. \cite{Sondhi97,Kvyatkovskii01,Vojta03,Belitz05,Sachdev00} for recent reviews).
The special case of structural phase transitions deserves, in our opinion, a special attention. First, it is very convenient when introducing the topic of low-$T$ phase transitions although, to the best of our knowledge, this pedagogical facet of the
structural case has not been developed in the literature. One of the purposes 
of the present paper is just to develop this facet. 
Second, the discussion of structural transitions allows to reveal
some unsolved problems which might have a fairly broad interest.

It is worth mentioning that our study will be restricted
to the region of small fluctuations (not very close to the phase-transition
point). This region normally is not the region of main interest in the
aforementioned papers, but the main specific features of the phase
transition anomalies are clearly seen already in this region, not to
mention that for interpretation of the experimental data this region is
quite often the most relevant one. 

A considerable part of the theory of low-$T$ structural phase
transitions is very simple. Its formulation uses elementary 
formulas of quantum and statistical mechanics, and its development involves a fairly simple
mathematics. Nevertheless, this elementary theory suffices to discuss some points of general interest
such as the validity of the Landau theory, the soft-mode concept, the role of quantum fluctuations
in defining the phase-transition point, the specific features of the
low-$T$ phase diagram, etc. This constitutes the first part of the
paper where, because of pedagogical considerations, we use a very
simple model. Nevertheless, even within this elementary treatment, there
arise some questions as well as not completely justified assumptions which
will be discussed in the second part of the paper.

These questions and assumptions refer to the character of the
dynamics of the order parameter near the zero- and low-$T$ phase
transitions. This character has not been successfully explained for high-$T$ 
phase transitions: the origin of the so-called central peak in the
soft mode spectrum is understood only qualitatively \cite{Ginzburg80}. For zero- and low-$T$ 
structural transitions this question has not been studied at all, although the dynamics of the order parameter is much more important here. Indeed, according to the
classical statistical mechanics the static properties of the system do not
depend on its dynamics. This is because (gaussian) integration over momenta
simply gives a factor in the corresponding partition function. But the
situation is different when quantum effects play a role. In this case, the
partition function does not factorize because momenta and coordinates, now
operators, do not commute with each other \cite{note1}. Therefore, a lack of exact knowledge
of the dynamics impedes obtaining definite results for, e.g., such a
``static'' property as dependence of the phase transition temperature on a
control parameter (e.g., strain or pressure) in the low-$T$ region. Given this
situation, we will discuss several possibilities without proposing
a finite conclusion about which of them corresponds to the reality. For this
discussion we need no model at all, and the system is considered in this
second part as a continuous medium.

\section{The single-ion model}

\begin{figure}[t]\centering
\includegraphics[width=.1\textwidth,clip]{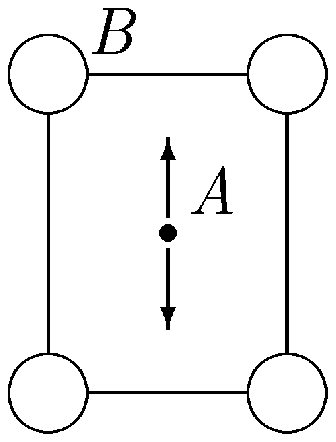}

\vspace{10pt}

\includegraphics[width=.175\textwidth,clip]{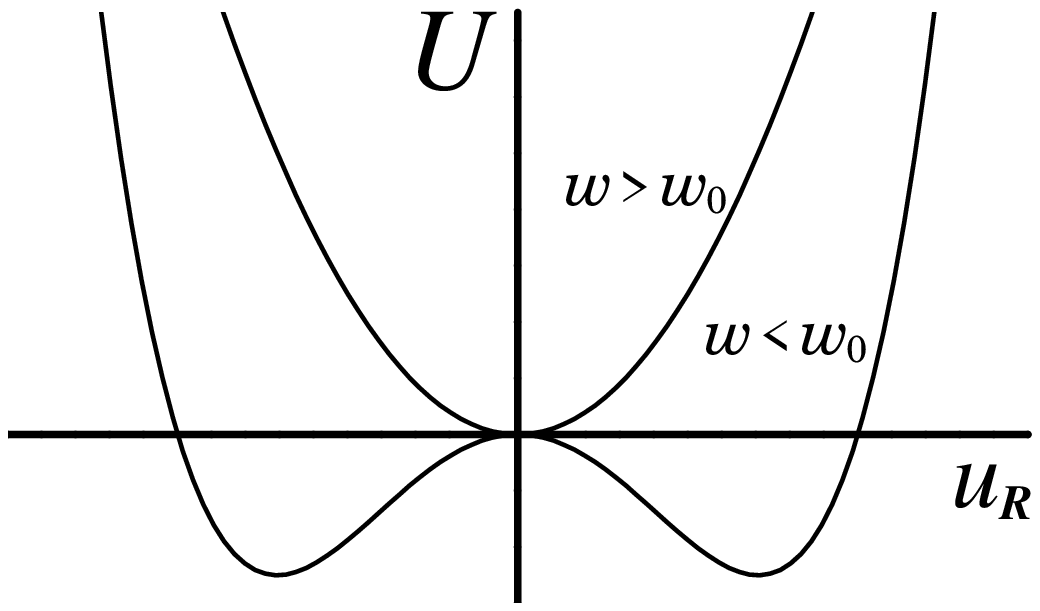}
\hspace{5pt}\includegraphics[width=.275\textwidth,clip]{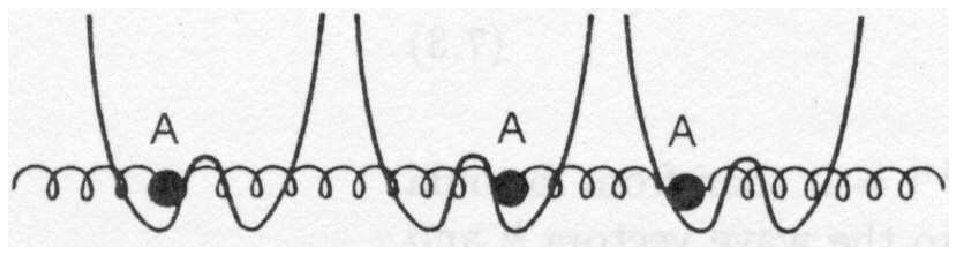}
\caption{The model: unit cell, effective potential acting on $A$-atoms due to $B$ ones, and an illustration of the interaction between $A$-atoms.}
\label{fig:1}\end{figure}

The so-called single-ion model (see, e.g., Ref. \cite{Strukov_Levanyuk}) is very convenient when illustrating a zero-$T$ structural phase transition. Within this model one assumes, first of all, that the crystal is composed by two types of atoms, say $A$ and $B$. Our aim is to describe ``active'' $A$-atoms in the simplest way, so we further assume that i) the sublattice of $B$-atoms can only be deformed homogeneously and ii) the interaction between $A$ atoms is a nearest-neighbor interaction mediated by springs. Additionally, there is an interaction between $A$ and $B$ atoms which is responsible for the relative position of the corresponding sublattices. Restricting ourselves to the orthorhombic case, let us choose the unit cell with $B$-atoms placed at the apices of the corresponding cell (see Fig. \ref{fig:1}). Thus, the potential acting on $A$-atoms due to the $B$ ones has to be symmetric with respect to the center of this cell. This is so if this potential has i) a minimum in the center of the unit cell or ii) two symmetric out-of-center minima. In the following we shall assume that i) is the case when the crystal is strongly compressed and then, along the $z$-axis, it turns into case ii) with diminishing the compression (see Fig. \ref{fig:1}). This makes possible a change in the mean position of $A$-atoms, i.e. a phase transition, in a fairly simple way. 

The potential energy of the system then can be written as 
\begin{align}
U= U_0 
+\sum_{\bm R}\left({a\over 2}u_{\bm R}^2 + {b\over 4}u_{\bm R}^4\right)
+\sideset{}{'}\sum_{\bm R,\bm R'}{c\over 2}
(u_{\bm R}-u_{\bm R'})^2,
\label{potential}\end{align}
where $u_{\bm R}$ represents the displacement of the $A$-atom along the $z$-axis in the $\bm R$th unit cell. The first sum in this expression represents the effective potential acting on $A$-atoms due to $B$ ones. Let us characterize the compression of the system by the magnitude $w= (V_0 - V)/V_0$, where $V $ is the volume of the system and $V_0$ is this volume at zero pressure for the (nonequilibrium) configuration in which all $A$-atoms are maintained in the center of the corresponding unit cells (i.e., $u_{\bm R}=0$). Thus, by taking $a=\alpha(w-w_0)$, with $\alpha>0$, and $b$ as a positive constant; $w_0$ gives the strain at which the form of this potential change from one-well to two-well. [The usually small difference between $V$ and $V_0$ ($|w|,|w_0|\ll 1$) turns out to be relevant for the change in the sign of the coefficient $a$ only, so we shall not distinguish between $V$ and $V_0$ anywhere but here.] The second sum in Eq. \eqref{potential} is the interaction potential between $A$-atoms, where $c$ is the stiffness coefficient of springs linking pairs of $A$-atoms (see Fig. \ref{fig:1}) and summation is carried out over nearest-neighbors only. 

\subsection{Static properties: A classical zero-$T$ transition}

Let us suppose at this point that the mass of $A$-atoms is infinite, so they can be treated as classical particles. Consequently, the configuration of the system will be the one which simply minimizes the potential energy. The static properties of the system will be in accordance with this configuration, so let us proceed to determine it. 

It is clear that the minimum of the potential energy corresponds to the configuration in which the
springs linking $A$-atoms do not experience any deformation. So all the atoms will be located in the same minimum of the effective potential created by $B$-atoms: $u_{\bm R}=u_0$. Eq. \eqref{potential} then reduces to
\begin{align}
U= U_0 + N \Big(
{a\over 2}u_{0}^2 + {b\over 4}u_{0}^4
\Big),
\label{}\end{align}
where $N$ is the number unit cells. Minimizing this potential we find the equilibrium value of $u_0$:
\begin{align}
u_{0,\text{eq}}^2=\begin{cases}
0& (w>w_0),\\
-{\displaystyle a\over \displaystyle b}&(w<w_0),
\end{cases}
\label{eq_u}\end{align}
and the corresponding value of the potential energy:
\begin{align}
U_{\text{eq}}= \begin{cases}
U_0&(w>w_0),\\
U_0 -N{\displaystyle a^2\over \displaystyle 4b}&(w<w_0).
\end{cases}\label{}\end{align}
In accordance with these formulas, the change in the form of the effective potential acting on $A$-atoms and the phase transition take place simultaneously at $w_0$. 

\subsubsection{Phase transition anomalies}

\begin{figure}[t]\centering
\includegraphics[width=.3\textwidth,clip]{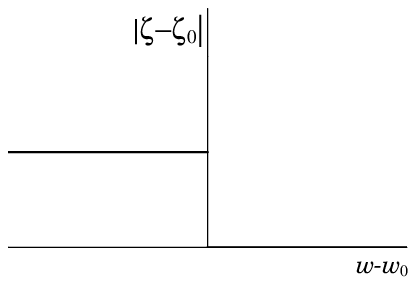}
\caption{\label{fig:AnomaliasLandau}}\end{figure}

As we have mentioned, the sublattice of $B$-atoms, and therefore the system as a whole, can be compressed homogeneously by applying pressure. Let us see how the corresponding stiffness of the system changes as a result of the phase transition. To this end, we present the potential energy of the system with $A$-atoms at the center of the corresponding unit cell as
\begin{align}
U_0\simeq U_{0}^\circ + {U_0 ' \over 2}w^2 + \dots 
\end{align}
(the term linear in $w$ is absent by virtue of the definition of $w$), and further take into account the energy of the mechanism applying the pressure: $-P (V_0 - V)=-PV_0w$. Minimizing the total energy with respect to $w$ we find this pressure:
\begin{align}
P= \begin{cases}
\zeta_0 w&(w>w_0),\\
\zeta_0 w -  n{\alpha^2 \over 2b} (w - w_0)  &(w<w_0),
\end{cases}\end{align}
where $\zeta_0 = U_0' /V $ and $n = N /V$. As a result, the stiffness is 
\begin{align}
\zeta = {dP\over dw }=
\begin{cases}
\zeta _0&(w>w_0),\\
\zeta_0  -  n{\alpha ^2 \over 2b}&(w<w_0).
\label{compressibility}\end{cases}
\end{align}
Its behavior is illustrated in Fig. \ref{fig:AnomaliasLandau}, which is just the standard behavior for an anomaly described within the Landau theory of second-order phase transitions.

\subsubsection{Exactness of the Landau theory for the model}

It is worth noticing that we have no fluctuations within this model, neither thermal (as long as $T=0$), nor quantum (as long as the masses are infinite). This makes Landau theory \cite{Landau37,Landau_SP,Strukov_Levanyuk} to be an exact theory here. It is frequently said that the Landau theory is a mean-field approximation, but this example shows that this statement is highly inappropriate. 

The possibility of having a phase transition with no fluctuations questions another frequent saying: ``zero-$T$ phase transitions are due to quantum fluctuations.'' Why? We had no fluctuations but we had a transition. Equally unjustified is to say that high-$T$ phase transitions are due
to thermal fluctuations. Thermal fluctuations give rise to some contribution, of course, but as a rule they are not the unique reason of the transition. The origin these sayings may be in the fact that phase transitions are frequently illustrated using the Ising model. For this model, it is true that the transition is due to the thermal fluctuations only. But while the Ising model played an important role in the theory of phase transitions, it is very specific, reflecting (as any model) no
more than an aspect of much a more many-faceted reality.

\subsection{Dynamics}

Let us now consider that the masses of $A$-atoms are finite, although very large; so that the motion of these atoms is possible, but the corresponding dynamics can still be considered as classical. Let us study some specific features of such a dynamics. To do so, it is convenient to introduce the new variables
\begin{align}
u_{\bm k} = {1\over N}\sum_{\bm R}u_{\bm R}e^{ - i \bm k \cdot \bm R}
\end{align}
having the meaning of the Fourier components of the displacement field:
\begin{align}
u_{\bm R} = \sum_{\bm k}u_{\bm k}e^{i \bm k \cdot \bm R}.
\end{align}

In terms of these variables the potential energy Eq. \eqref{potential} has the form
\begin{align}
U &= U_0 + N\sum_{\bm k} {a(\bm k)\over 2}|u_{\bm k}|^2  
\nonumber \\ &\quad +N{b\over 4}\sum_{\bm k, \bm k ' , \bm k '' } u_{\bm k} u_{\bm k'}u_{\bm k''}u_{- \bm k - \bm k' - \bm k''},
\end{align}
where $a(\bm k)=a + 4c (\sin ^2 {k_x l \over 2} +\sin ^2 {k_y l \over 2} + \sin ^2 {k_z l \over 2})$, with $l$ being the cell parameter (here it has been taken into account that $u_{-\bm k}= u_{\bm k}^*$). By separating the zero Fourier component of the displacement field: $u_0=N^{-1}\sum _{\bm R} u_{\bm R}$, which evidently has the meaning of the mean value of the displacement field, the above expression for the potential energy can be rewritten as
\begin{align}
U &= U_0 + N \Big( {a\over 2}u_0^2 + {b\over 4}u_0^4 \Big)
+ N \sum_{\bm k \not =0} {a(\bm k, u_0)\over 2}|u_{\bm k}|^2  
\nonumber \\&\quad 
+ N{b}u_0 \sum_{\bm k, \bm k ' \not =0} u_{\bm k} u_{\bm k'}u_{- \bm k - \bm k' }
\nonumber \\&\quad 
+N{ b\over 4}\sum_{\bm k, \bm k ' , \bm k '' \not =0} u_{\bm k} u_{\bm k'}u_{\bm k''}u_{- \bm k - \bm k' - \bm k''}.
\label{Full_U}\end{align}
where $a(\bm k,u_0) = a(\bm k)  + 3bu_0^2$. Below we shall consider the case of small-amplitude oscillations such that the contribution of the last two sums in this expression can be neglected. It is worth noticing that this does not mean to neglect all the anharmonicity of the system as long as the coefficient $b$ is still present in the remaining terms: 
\begin{align}
U &\simeq U_0 + N \Big({a\over 2}u_0^2 + {b\over 4}u_0^4 \Big)
+N\sum_{\bm k \not =0} {a(\bm k, u_0)\over 2}|u_{\bm k}|^2  .
\label{U_WeakAnharmonicity}
\end{align}

\begin{figure}[t]\centering
\includegraphics[width=.3\textwidth,clip]{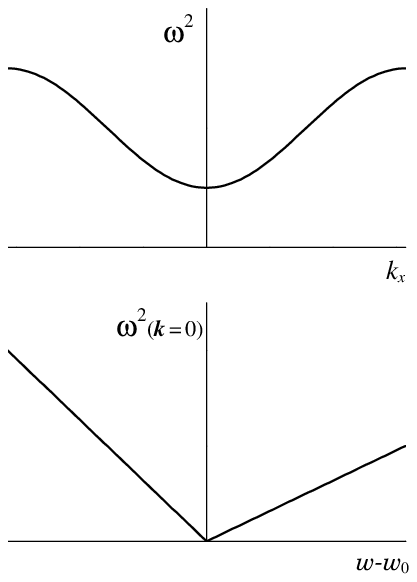}
\caption{\label{fig:Dispersion}}\end{figure}

In the simplest case, the equation of motion for the Fourier components of the displacement field can be written as 
\begin{align}
m \ddot u_{\bm k} + a(\bm k,u_0) u_{\bm k} =0,
\label{motion_phonon}\end{align}
where $m$ is the mass of the $A$-atoms. We then have an optical branch with the dispersion law $\omega_c^2(\bm k, u_0) = a(\bm k,u_0)/ m $ (see Fig. \ref{fig:Dispersion}). For small wavevectors: 
\begin{align}
\omega_c^2(\bm k, u_0) = (a + 3b u_0^2 + \widetilde c k ^2)/m,
\label{omega_c}\end{align}
where $\widetilde c = c l^2$. Substituting here the equilibrium
value of $u_0$, Eq. \eqref{eq_u}, we obtain the normal frequencies of the system as a function of the control parameter $w$. For $\bm k = 0$ that is
\begin{align}
\omega_c^2 (0, u_{0,\text{eq}}) =
\begin{cases}
{\alpha |w - w_0|\over m}& (w>w_0),\\
{2\alpha |w - w_0|\over m}&  (w<w_0).\\
\end{cases}\end{align}
This behavior is illustrated in Fig. \ref{fig:Dispersion}. It is worth noticing that, at $w = w_0$, this optic branch has the same dispersion law as the acoustic one: $\omega \propto k$. As a result of this behavior, first noticed by Ginzburg \cite{Ginzburg49}, this $\bm k = 0$ mode is usually termed as the soft mode associated with the transition.

\subsection{Zero-$\bm T$ transition: Quantum effects}

Let us now consider that the masses of $A$-atoms are ``normal'' (not specially large). In this case, if we consider $A$-atoms as immobile atoms when studying the thermodynamic properties of the system we are wrong. Even at $T=0$ these atoms are moving exhibiting what is called quantum fluctuations, and we have to take into account this quantum effects. 

In a first step, the system can be considered as a set of harmonic oscillators [see Eq. \eqref{motion_phonon}]. For a fixed value of $u_0$, the ground state energy of the system then can be written as
\begin{align}
E = U_0 +  N \Big({a \over 2}u_0^2 + {b\over 4}u_0^4\Big) +\sum _{\bm k} {\hbar \omega_c(\bm k, u_0)\over 2}
\label{FreeEnergy_Phonons}\end{align}
(recall that $T=0$).
As a result of the possible displacements of $A$-atoms along the $z$-axis, here we have the classical (macroscopic) contribution considered in previous section plus a quantum term accounting for the ground-state energy of the corresponding oscillators with normal frequencies $\omega_c(\bm k, u_0)$. In further calculations it is convenient to take the continuous-medium limit of our model (i.e., to replace summation by integration $\sum_{\bm k} \to V \int {d\bm k\over (2 \pi )^3}$): 
\begin{align}
E = U_0 +  N \Big({a \over 2}u_0^2 + {b\over 4}u_0^4\Big) + V\int{d \bm k\over (2\pi)^3}{\hbar \omega_c(\bm k, u_0)\over 2}. 
\label{FreeEnergy_PhononsContinuous}\end{align}

\subsubsection{The phase-transition point}

Because of the dependence on $u_0$ of the quantum term ($\propto \hbar$) in Eq. \eqref{FreeEnergy_PhononsContinuous}, it can be said that there effectively is a quantum renormalization of the function $E(u_0)$. By expanding the last term in Eq. \eqref{FreeEnergy_PhononsContinuous} in power series of $u_0$ we can obtain, for example, the quantum correction to the coefficient $a$. The resulting coefficient, $a^*$, is of particular interest to us: when $a^*=0$ the system losses its stability with respect to nonvanishing values of $u_0$, what just defines the phase-transition point. Taking into account that 
\begin{subequations}
\begin{gather}
\left.{\partial \omega_c (\bm k,u_0)\over \partial u_0}\right|_{u_0=0}=0,\\
\left.{\partial^2 \omega_c (\bm k,u_0)\over \partial u_0^2}\right|_{u_0=0}={3b\over m\omega_c(\bm k,0)},
\end{gather}\end{subequations}
we find that
\begin{align}
a^* = a  + {3 \hbar b v\over 2 m}\int {d \bm k\over (2\pi)^3}{1\over \omega_c(\bm k,0)}.
\label{renormalized_a}\end{align}
where $v=V/N=l^{3}$ is the volume of the unit cell. When trying to determine the phase-transition point one realizes that this formula is, however, somewhat contradictory. The coefficient $a$ has to be negative in order to further obtain $a^{\ast }=0$. But in this case, in accordance with Eq. \eqref{omega_c}, there is a region of wavevectors for which $\omega_{c}^{2}< 0$.

This difficulty can be circumvented if this region is much smaller than the whole Brillouin zone.
When this possibility takes place the system can be labeled as displacive and, when calculating the integral in Eq. \eqref{renormalized_a}, one can put $a=0$ quite safely. As a result one finds that the phase-transition strain is given by $w_{c}=w_{0}+\delta w$, where $\delta w$ is the contribution due to quantum effects (zero-point quantum fluctuations) which can be estimated as 
\begin{align}
\delta w \simeq - {3\hbar b v\over  (2 \pi)^2 \alpha m^{1/2}}\int_0^{2\pi/l} {k^2 d k\over  \sqrt{a + \widetilde c k^2} }
\simeq - {3 \hbar b \over 2 \alpha \sqrt{m c}}.
\end{align}
Alternatively, by realizing that the frequency of the soft mode should vanish at the
``experimental'' value of phase transition strain $w_c$, one could replace $a \to a^{\ast }$ in
the integral in Eq. \eqref{renormalized_a}. This is equivalent to take into account, in an effective way, higher order corrections in Eq. \eqref{renormalized_a} (see below). 

Although the problem seems to be overcame, it is instructive to discuss this ``overcoming'' in more detail. What we have done is to ``correct'' the dynamical moduli of the system in a meaningful way. This is quite similar to what is done when calculating, e.g., the phase transition temperature of a classical displacive system: here they are the static moduli what are corrected in a proper way. By realizing that our system is in fact an anharmonic system, as indicated by the last
two sums in Eq. \eqref{Full_U}, it can be expected that going beyond the approximation of decoupled oscillators such a correction (renormalization) will be obtained. The possibility of using a perturbation theory to account for this anharmonicity is therefore quite attractive, and valuable irrespective of the character, classical or quantum, of the problem in question. However, the quantum case is somewhat more complex than the classic one as long as dynamics plays a key role. One can realize that we did two things in fact: i) assumed that dynamics is still well-described in the same terms (i.e., within a soft-mode scenario) and then ii) corrected the corresponding dynamical moduli. The former has not to be necessarily true in a real case. Indeed, one of the well known shortcomings of the theory of high-$T$ structural phase transitions is that the soft mode behavior has never been observed experimentally to a full extent: the ``soft mode'', at best, diminishes its frequency considerably; but this frequency never goes to zero at the phase transition point. At the same time, close to the phase transition point there appears the so-called ``central peak'' in the order parameter fluctuation spectrum, and the increase of the order parameter fluctuations turns out to be comprised within this peak. Such a peak is a natural feature of the so-called order-disorder systems, for which the order-parameter dynamics is relaxational instead of phonon-like as we have considered so far. Though there is no consistent analytical theory of the central peak, its appearance is not surprising at all \cite{Ginzburg80}. Any system with a phase transition is inherently anharmonic. So the phonon-like (``soft mode'') picture is not the full picture; there is another part, essentially an anharmonic one, which reveals itself in the central peak. Similar effects are quite possible at low $T$s: the presentation of the motion as a set of normal vibrations, even followed by taking into account the anharmonism within a perturbation theory, may fail to describe accurately the dynamics of the system close to the corresponding phase-transition point. Consequently, one can do nothing but assume Eq. \eqref{renormalized_a} with caution and hope that it is qualitatively correct.

The importance of the quantum corrections to phase-transition strain can be estimated as follows. The very possibility of observing the transition implies that $|w_0|$ is much less than the atomic value of this magnitude $|w_\text{at}| \sim 1$. The other constants in the model, however, can have their ``atomic'' values. Introducing the atomic (binding) energy $\varepsilon_\text{at}\sim \hbar^2/(m_e l^2)$, where $m_e$ is the electron mass and $l\sim 1 \text{\AA}$ the atomic distance, we then have 
\begin{align}
\alpha \sim {\varepsilon_\text{at}\over l^2 w_\text{at}},\quad b\sim {\varepsilon_\text{at}\over l^4},\quad c\sim {\varepsilon_\text{at}\over l^2}. 
\label{ForTheEstimates}\end{align}
Thus we find that 
\begin{align}
\left|{\delta w \over w_0}\right|\sim \left({m_e\over m}\right)^{1/2} \left|{w_\text{at}\over w_0}\right| .
\end{align}
As long as $m_e \ll m$, but ${|w_\text{at}|\gg |w_0|}$, here we see that the correction to the classical phase-transition strain can be quite important: it may give rise to a magnitude $w_c$ comparable, or even greater, than $w_0$.

\subsubsection{Order-disorder (spin-like) limit}

We have discussed the displacive limit but what happens if the value of $|a|$ at the phase-transition point is so large that $\omega_{c}^2<0$ in most of the Brillouin zone? In this case, one simply has to realize that the starting point was not chosen
appropriately. Looking back we can see that this case corresponds to an effective potential acting on $A$-atoms due to $B$ ones with two profound wells even in the symmetric phase. Under these circumstances, instead of characterizing the motion of $A$-atoms by their displacement from the center of the corresponding unit cell (what $u_{\bm R}$ really means), it makes more sense to assume, first of all, the atoms to be confined inside one of the two wells and then take into account the possibility of delocallization by, e.g., quantum tunneling. This can be done by associating the two initial positions of $A$-atoms with the two possible orientations of (pseudo-)spins $1/2$ and accounting for the presence of a transversal field in order to reproduce the tunneling. Thus our model becomes a spin model exhibiting a zero-$T$ phase transition (Ref. \cite{Sachdev00,Vojta03}). The corresponding excitations (spin waves) are similar to the optical vibrations considered above in the sense that there is a ``soft spin-wave'' whose frequency goes to zero at the phase-transition point. In this sense, when it comes to zero-$T$ transitions, displacive and order-disorder limiting cases are not so different one another.

\subsubsection{Phase-transition anomalies}

Let us now calculate the quantum contribution to the anomaly in the stiffness of the system within the displacive scenario. The main purpose of such a calculation is to estimate the region in which the Landau theory is applicable, so it suffices to consider the symmetric phase ($u_{0}=0$). From Eqs. \eqref{FreeEnergy_Phonons} and \eqref{omega_c}, we find that 
\begin{align}
\zeta &= {1 \over V}{\partial ^2 F \over \partial w^2 }= 
\zeta _0   -{\hbar \alpha^2 \over 8 m^2} 
\int {d{\bm k } \over (2\pi )^3}{1 \over \omega_c^{3/2} (\bm k,0)}
\nonumber \\&
\simeq \zeta_0 - {\hbar n \alpha^2  \over 32 \pi^2 m^{1/2} c^{3/2}} 
\ln \left( { 4  \pi ^2 c \over \alpha (w-w_0)}\right),
\label{compressibility_quantum_corrections}\end{align}
where $\zeta_0$ is in accordance with Eq. \eqref{compressibility}. Let us analyze this result in some detail. 

At first sight, it seems that the stiffness of the system changes its sign
and diverges at $w=w_{0}$. However, as long as we are considering the lowest
order correction only, we simply can say that it diminishes close to $w_{0}$. 
In any case, why close to $w_0?$ An anomaly at the phase-transition point would be not
surprising but, in accordance with Eq. \eqref{renormalized_a}, the
phase-transition point is not $w_{0}$. What happens? The
answer is that here have the same kind of inconsistency that we have found before [compare the integral in Eqs. \eqref{renormalized_a} and \eqref{compressibility_quantum_corrections}]. So the replacement of $w_0$ by the actual value associated with the phase-transition strain, which can be extracted from the experiments, solves it.

An estimate of the region of applicability of the Landau theory in our case
can be made by demanding that the (quantum) contribution to the stiffness of
the system is much smaller than the discontinuity obtained within the Landau
theory itself. That is 
\begin{align}
{\hbar  n \alpha^2 \over 32 \pi^2 m^{1/2}c^{3/2}}
\left| \ln \left( { 4  \pi ^2 c \over \alpha (w-w_c)}\right) \right| \ll n{\alpha ^2 \over 2 b}
\label{}\end{align}
[see Eq. \eqref{compressibility_quantum_corrections}]. In accordance with previous estimates of the coefficients appearing in this expression [see Eq. \eqref{ForTheEstimates}], this gives
\begin{align}
\left|\ln \left( { w - w_c\over w_\text{at}} \right)\right| \ll 10^2  \left( {m \over m_e} \right)^{1/2}\sim 10^4.
\end{align}
As wee see, it is hardly possible to abandon this region in real experiments. 

\begin{figure}[t]\centering
\includegraphics[width=.3\textwidth,clip]{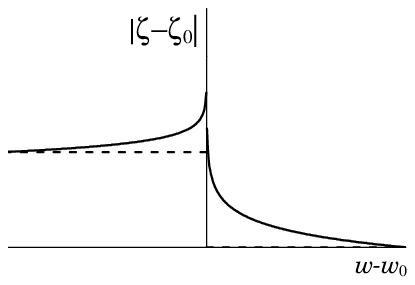}
\caption{\label{AnomaliasPrimerOrden}}\end{figure}

It is worth mentioning that the behavior of the stiffness of our model in the scaling region has been known since long ago \cite{Rechester71}. Roughly speaking, it is such that one has to replace $\ln $
by $\ln ^{1/3}$ in Eq. \eqref{compressibility_quantum_corrections}. As we see, both the first-order approximation and the theory for the scaling region provide qualitatively the same results (see Fig. \ref{AnomaliasPrimerOrden}).
However, as a rule, the first-order approximation gives rise to an overestimate of the relevant magnitudes in the very vicinity of the phase-transition point: in our case, by extrapolating the increase of the stiffness obtained within the first-order approximation to the scaling region one overestimates such an increase. 

\subsubsection{Thermal effects}

Thermal effects can be revealed by following the same considerations as above. 
At finite temperature, the free energy of the system can be written as
\begin{align}
F &= F_0 +  N \Big({a \over 2}u_0^2 + {b\over 4}u_0^4\Big) + V\int {d\bm k\over (2\pi)^3}{\hbar \omega_c(\bm k, u_0)\over 2} 
\nonumber \\&\quad 
+ {T }V\int {d\bm k\over (2\pi)^3}\ln \left\{ 1 - \exp[- \hbar \omega_c(\bm k, u_0)/T] \right\}.
\label{FreeEnergy_PhononsT}\end{align}
(Notice that this expression is nothing but the free energy of a set of harmonic oscillators, see, e.g., Ref. \cite{Landau_SP}.) Therefore, the coefficient at the term quadratic in $u_0$ is 
\begin{align}
a^* &= a  + {3 \hbar b v\over 2 m}\int {d \bm k\over (2\pi)^3}{1\over \omega_c(\bm k,0)} 
\nonumber \\ &\quad 
+ {3 \hbar b v\over 4 m}\int {d \bm k\over (2\pi)^3}{n [\omega_c(\bm k,0)] \over \omega_c(\bm k,0)},
\label{renormalized_aT}\end{align}
where $n(\omega) = [\exp(\hbar \omega /T) - 1]^{-1}$ is the Bose-Einstein distribution function. Here we have the quantum contribution computed in previous section [see Eq. \eqref{renormalized_a}] plus a thermal one (the latter integral). Similar to what we have done when discussing the quantum
contribution to $w_{c}$, it is reasonable to replace $a \to a^{\ast }$ 
in the expression of $\omega _{c}$ entering in Eq. \eqref{renormalized_a}. 

Let us discuss the phase diagram. The border between the symmetric and non-symmetric phases in the ($T,w$)-plane is defined by the condition $a^{\ast }=0$. This gives the line
\begin{align}
 w-w_{c} =-\alpha^{-1} f\left(T_c\right),
\end{align}
in the ($T,w$)-plane, where $f\left( T_c\right) $ is the last integral in Eq. \eqref{renormalized_aT} evaluated at $a^*=0$. At low temperatures [$\hbar \omega(0,0)\ll T \ll \hbar \omega(\bm k_\text{max},0)$], this integral can be estimated as 
\begin{align}
&\int {d \bm k\over (2\pi)^3}{n [\omega_c(\bm k,0)] \over \omega_c(\bm k,0)}
={1\over 2\pi^2}\int {n[\omega_c(k,0)]\over \omega_c(k,0)}k^2 d k
\nonumber \\&\quad\simeq 
{1\over 2\pi ^2}\int_0^{k_\text{max}(T)} {T\over \hbar \omega_c^2(k,0)}k^2dk
={m^{3/2}\over 2\pi^2 \hbar ^2 c^{3/2}v}T^2.
\end{align}
taking into account that the Bose-Einstein distribution function decreases very strongly within the integration interval [it vanishes for $k \gtrsim k_\text{max}(T)= {m^{1/2} \over \hbar \widetilde c^{1/2}}T$]. This gives
\begin{align}
w-w_c \propto T^2_c.
\label{}\end{align}

At high temperatures [$ \hbar \omega_c(\bm k,0)\ll T $], the Bose-Einstein distribution can safely be replaced by its classical limit:
\begin{align}
\int {d \bm k\over (2\pi)^3}{n [\omega_c(\bm k,0)] \over \omega_c(\bm k,0)}\simeq 
\int {k^2 d k\over 2\pi^2}{T \over \hbar \omega_c^2(\bm k,0)}\simeq 
{m \over \pi \hbar c v}T.
\label{}\end{align} We then have
\begin{align}
w-w_c\propto T_c.
\label{}\end{align}

\section{Continuous medium approach}

At this point, it is worth making a comparison between origin of the quantum and thermal contributions in $a^*$ obtained previously [see Eqs. \eqref{renormalized_a} and \eqref{renormalized_aT}]. While the quantum contribution comes from the
ground-state energy of all the optical phonons, the thermal
contribution is obtained from the optical phonons with small wavevectors
only. In consequence, the actual value of phase-transition strain at zero
temperature ($w_{c}$) is sensitive to the microscopic details of the system,
whereas the function $T_{c}(w)$ defining the phase transition temperature as
a function of strain is not. This can be taken as a justification
to use a continuous medium approach when studying thermal dependencies. Just in this sense one speaks about an universal low-$T$ behavior of the systems, exemplified by
the temperature dependence of specific heat of solids (the Debye law): it
comes from the contribution of small-wavevector acoustic phonons and,
therefore, is characterized by macroscopic quantities such as the velocity
of sound. 

However, in our case, this universality is not so evident. It can be expected when the soft-mode scenario takes place, as we were able to argue for a classical zero-$T$ transition. 
But as soon as we make the model more realistic, taking
into account real values of the ion masses and, therefore, the quantum
effects, as well as considering other degrees of freedom, in particular,
acoustic modes, the dynamics of the system becomes far more complicated. This
has not been studied to a full extent, for which there is a reasonable explanation: difficulties similar to those found for the high-$T$ order-parameter dynamics are indeed expected. 
As we have mentioned, the soft-mode scenario is not completely applicable to high-$T$ phase transitions: in both real and numerical experiments the central-peak phenomenon appears time and again, and this lacks a theoretical explanation beyond ``hand-waving'' arguments \cite{Ginzburg80}. 
The situation for zero- and low-$T$ phase transitions is unknown, but one may suspect that it is similar. In consequence it is quite reasonable to consider, within the continuous medium approach, both hypothetical cases: that of the soft-mode (phonon-like dynamics) scenario and the
central-peak-like (relaxational dynamics) one. As we shall see, the aforementioned universality is lost in the second case for which the continuous medium approach becomes no more than a model.

\subsection{Phonon-like dynamics\label{S:Phonon-like dynamics}}

Our arguments about the universality were referred just for the phonon-like
case. In this case, the theory can be made somewhat more realistic, even model-independent, than the one presented above where only one (``active'') optical branch is taken into account. It is straightforward, in particular, to take into account the acoustic branches (e.g., to account for the motion of $B$-atoms as well). Indeed one may expect that these acoustic phonons give rise to a significant contribution to $a^{\ast }$: they are low-frequency excitations of the system. Consider the free energy 
\begin{align}
F &= \widetilde F_0 +  N \Big({a^*(T)\over 2}u_0^2 + {b\over 4}u_0^4\Big) 
+ 3V\int {d\bm k\over (2\pi)^3}{\hbar \omega_\text{ac}(k, u_0)\over 2} 
\nonumber \\&\quad 
+ 3{T }V\int {d\bm k\over (2\pi)^3}\ln \left\{ 1 - \exp[- \hbar \omega_\text{ac}(k, u_0)/T] \right\}.\label{}
\end{align}
where $a^*(T)$ is given by Eq. \eqref{renormalized_aT}. This is nothing but Eq. \eqref{FreeEnergy_PhononsT} where the contribution of the acoustic phonons has been added explicitly. This contribution depends on $u_0$ because of the corresponding dependence of the velocity of sound:
\begin{align}
\omega_\text{ac}(k, u_0)= c(u_0)k,
\end{align}
which can be taken as $c(u_0)= c_0(1 + {\alpha \over 2}u_0 ^2)$. (This velocity has to be understood as an averaged velocity of sound in the same sense as in Ref. \cite{Landau_SP}.) Thus the new coefficient of the free energy at the quadratic term in $u_0$ can be written as
\begin{align}
a^{**}(T)&= a^*(T) \nonumber \\
&\quad + {3\hbar \alpha v \over 2 \pi^2} \int 
\left\{ {1\over 2} + n [\omega_\text{ac}(k,0)] \right\} \omega_\text{ac}(k,0){ k^2 d k}.
\label{renormalized_aT_acoustic}
\end{align}
As we see in this expression, we have a new contribution to the phase-transition strain $w_c$ which is not surprising (the part of the integral which does not vanish at $T=0$). But we also have a thermal one which can be estimated as
\begin{align}
{3\hbar \alpha v\over 2\pi^2 } \int {n[\omega_\text{ac}(k,0)]\omega_\text{ac}(k,0)}k^2 d k
\simeq 
{\hbar c_0\alpha v\over 2\pi^2 } \left( {T\over \hbar c_0}\right)^4.
\end{align}
Having in mind that the thermal activation of the optical phonons obtained above is exponentially suppressed with moving away the phase-transition point [see Eq. \eqref{renormalized_aT}], one realizes that this thermal contribution due to acoustic phonons may be the most important one in a significant region of the ($T,w$)-plane. However, this is not the case close enough to the phase transition point where $a^*(T)- a^*(0) \propto T^2$. Consequently, the form of the phase diagram does not change from the one expounded above.

These changes take place if, for instance, there exist long-range interactions in the
system. In this case, complicated dispersion relations may arise such that the dimensionality of the integrals in Eq. \eqref{renormalized_aT} effectively increase. In the case of uniaxial
ferroelectrics, for instance, dipolar interactions lead to $\omega_c^2(\bm k, 0)= (a + \widetilde c k^2 + 4\pi \cos^2 \theta)/m$ for small wavevectors. This further gives $a^*(T)- a^*(0) \propto T^3$ (see, e.g., Refs. \cite
{Khmelnitskii71,Kvyatkovskii01}). The linear coupling between the order parameter and one of the component of the strain tensor, i.e., the piezoelectric effect in the case of ferroelectrics, gives rise a peculiar dependence $a^*(T)- a^*(0) \propto T^{5/2}$ \cite{Cano04b}. These long-range interactions, however, do not modify the high-$T$ behavior.

\subsection{Dissipative dynamics}

Let us now consider that there is a central peak in the order-parameter
fluctuation spectrum, at least close to the phase-transition point. Such an
assumption is far from being trivial: it implies that there is an essential frequency dispersion in the order-parameter response function and, therefore, a dissipative dynamics of
the order parameter even at $T=0$. In a number of the systems, e.g., for perfect crystals without phase transitions, this is certainly not the case. But
the situation with such an inherently anharmonic system as the one we are considering here 
(recall that $a<0$ in our case) is unclear. What seems to be clear,
however, is that for crystals with some defects the dissipative dynamics of the
type we are supposing here takes place even at $T=0$ \cite{Cano04a}. Therefore, even if our
consideration would prove to be irrelevant for pure systems, it might still
be relevant for systems with some defects. A dissipative dynamics for a set
of variables means that these variables are not coordinates of a closed
system but there is a ``reservoir'' where the system in question transfers
its energy to. In our case, the long-wave acoustic phonons could play the role of such a reservoir. When calculating the contribution of the set of dissipative variables one should, of course, not forget the contribution of the reservoir. Only in the case when the latter contribution is
irrelevant it makes sense to consider a system of dissipative variables as
the main contributor (see, e.g., Ref. \cite{Weiss}). Let us mention that even apart from possibility of the central peak, an account of a phonon damping at $T=0$ due to e.g. defects is of some interest for zero- and low-$T$ transitions. We shall begin with this case by postulating the equation of motion 
\begin{align}
m \ddot u_{\bm k} + \widetilde \gamma \dot u_{\bm k} + a(\bm k,u_0) u_{\bm k} =0,
\label{dissipative_motion}\end{align}
where the viscosity coefficient $\widetilde{\gamma}$ accounts for the
dissipation.

\subsubsection{Thermal effects}

Making use of some results obtained by previous authors (see, e.g., Ref. \cite{Weiss}), we can study the case in which the motion of the ``active''-atoms is governed by Eq. \eqref{dissipative_motion}. In this case, the free energy can be written as \cite{Weiss}
\begin{widetext}\begin{align}
F &= \widetilde F_0 + N \Big({a \over 2}u_0^2 + {b\over 4}u_0^4 \Big)
+
{T}V\int {d\bm k\over (2\pi)^3}\left[\ln \Big({2 \pi [\lambda_1(\bm k, u_0)\lambda_2(\bm k, u_0)]^{1/2}\over \nu}\Big)
-\ln \Gamma\Big(1 + {\lambda_1(\bm k, u_0)\over \nu}\Big)-\ln \Gamma\Big(1 + {\lambda_2(\bm k, u_0)\over \nu}\Big)
\right],
\label{FreeEnergy_DissipativeT}\end{align}\end{widetext}
where $\nu = 2 \pi T/\hbar$ and $\{\lambda_i(\bm k, u_0)\}$ are the roots of the equation $\lambda ^2(\bm k, u_0) - \gamma \lambda(\bm k, u_0) + \omega_c^2(\bm k, u_0) =0$, satisfying the relations
\begin{subequations}\begin{gather}
\lambda_1 (\bm k, u_0)+ \lambda_2 (\bm k, u_0)= \gamma, \\
\lambda_1 (\bm k, u_0)\lambda_2(\bm k, u_0) = \omega_c ^2(\bm k, u_0).
\label{}\end{gather}\end{subequations}
It is worth mentioning that the first term $\widetilde F_0$ in Eq. \eqref{FreeEnergy_DissipativeT} now includes the contribution due to the reservoir. In principle, this contribution could exhibit a nontrivial behavior close to the transition point as a result of the coupling between the active atoms (which clearly ``feel'' the transition) and the degrees of freedom in the reservoir (which also ``feel'' the transition because of this coupling). Nevertheless, if these degrees of freedom forming the reservoir are low-frequency acoustic phonons, which would be quite natural, it can be shown that their contribution i) is not substantially modified because of damping \cite{Cano04a} and ii) is not the most important one close to the (low-$T$) phase-transition point as we have seen in Sec. \ref{S:Phonon-like dynamics} \cite{Cano04b}.

The above formulas allow us to compute the relevant magnitudes close to the phase transition by following the same procedure as before (see Ref. \cite{Cano04b}). The coefficient at the quadratic term in $u_0$ in the free energy, for instance, is found to be 
\begin{align}
a^*(T) &= a + 
{3b v\over m k_BT}
\int {d{\bm k} \over (2 \pi)^3}\bigg({1\over 2\omega_c^2(\bm k,0)}
\nonumber \\ &\quad 
+
{\psi[1+\lambda_1(\bm k,0)/\nu]-\psi[1+\lambda_2(\bm k,0)/\nu]\over \nu [\lambda_1(\bm k,0) - \lambda_2(\bm k,0)]}
\bigg).
\label{chi_simple}\end{align}
Close to the phase-transition point, this further gives
\begin{align}
a^*(T) - a^*(0)\propto T^2,
\label{}\end{align}
in the relaxation limit ($m\to 0$). It is worth mentioning that this behavior is found irrespective of long-range forces \cite{Cano04b}, which is related to the fact that the thermal activation of the elementary excitations of the system is substantially modified as a result of the damping [notice that the Bose-Einstein distribution function in Eq. \eqref{renormalized_aT} has been replaced by the psi functions in Eq. \eqref{chi_simple}]. In fact, the most important conclusion here is that, as a result of damping, the paradigm: low-$T$ behaviors $=$ small wavevectors, is not longer valid. This is because all degrees of freedom may give a contribution to the corresponding thermal behavior as a result of the damping [if $m\to 0$, in Eq. \eqref{chi_simple} there is no temperature-dependent cutoff analogous to that in Eq. \eqref{renormalized_aT}]. In some sense, this is similar to what happens in the case of the phase transition strain: one has to go beyond the continuous media approach to calculate this strain. The situation is new, however, in the sense that this happens even for the thermal behavior.

\section{Conclusions}

The subject of low-temperature phase transitions has been examined, dealing
with the structural case and focusing on the region of small fluctuations.
Within an oversimplified but illustrative model we have discussed some
points of general interest, such as the validity of the Landau theory, the
role of quantum fluctuations in defining the phase-transition point and the
specific features of the low-temperature phase diagram. We have pointed out also that a profound study of dynamics, both experimental and theoretical, is still needed to get fully justified results about the phase diagrams and anomalies at structural and may be other low-temperature phase
transitions.

\begin{acknowledgments}

We thank C. Barcel\' o and F.F. L\' opez-Ruiz for helpful comments on the manuscript.

\end{acknowledgments}


\begin{thebibliography}{0}
\bibitem{note1} This difference is even more apparent within the path integral formalism:
the domain of integration over the (imaginary) time variable turns out to be
infinitesimal in the classical case (high-$T$s), while it is not in
the quantum one (low-$T$s).
\end{thebibliography}
\end{document}